\documentclass{article}

\usepackage{graphicx}
\usepackage{amsmath}
\usepackage{amssymb}
\usepackage{amsfonts}

\usepackage{authblk}

\title{Singular effect of linkage on long term genetic gain in the infinitesimal model}

\author[1]{Elise Tourrette}
\author[2,3]{Olivier C. Martin}
\affil[1]{Universit\'e Paris-Saclay, INRAE, CNRS, AgroParisTech, GQE - Le Moulon, 91190 Gif‐sur‐Yvette, France}
\affil[2]{Universit\'e Paris-Saclay, INRAE, CNRS, Univ. Evry, Institute of Plant Sciences Paris-Saclay (IPS2), 91405 Orsay, France}
\affil[3]{Universit\'e Paris-Cit\'e, CNRS, INRAE, Institute of Plant Sciences Paris-Saclay (IPS2), 91405 Orsay, France}

\begin{document} 

\maketitle

\begin{abstract}
During the founding of the field of quantitative genetics, Fisher formulated in 1918 his ``infinitesimal model'' that provided a novel mathematical framework to describe the Mendelian transmission of quantitative traits. If the infinitely many genes in that model are assumed to segregate independently during reproduction, corresponding to having no linkage, directional selection asymptotically leads to a constant genetic gain at each generation. In reality, genes are subject to strong linkage because they lie on chromosomes and thus segregate in a correlated way. Various approximations have been used in the past to study that more realistic case of the infinitesimal model with the expectation that the asymptotic gain per generation is modestly decreased. To treat this system even in the strong linkage limit, we take the genes to lie on continuous chromosomes. Surprisingly, the consequences of genetic linkage are in fact rather singular, changing the \textit{nature} of the long-term gain per generation: the asymptotic gain vanishes rather than being simply decreased. Nevertheless, the per-generation gain tends to zero sufficiently slowly for the total gain, accumulated over generations, to be unbounded.
\end{abstract}


\vfill
\noindent
\textit{keywords:} recurrent selection | genetic value | genetic linkage  \\
\textit{abbreviations:} LD, linkage disequilibrium | CO, crossover | DS, Directional Selection | IM, infinitesimal model | i.i.d, independent and identically distributed  \\

\vfill
\noindent

In the early twentieth century, Fisher's infinitesimal model, in which each gene contributes an infinitesimal amount to a trait, re-conciliated the Mendelian and biometrical approaches for describing inheritance of continuous traits. Furthermore, it later provided justification for the steady rise of genetic values in directional selection programs. However, predictions for long-term behavior in that model have ignored genetic linkage. Surprisingly, we find that genetic linkage changes the model's behavior in a singular way, driving the system towards an ``aging'' regime in which the genetic gain per generation decreases down to 0. Nevertheless, because this slow down is so progressive, the gain that can be accumulated over generations is unlimited.



\section{Introduction}
At the end of the 19'th century, extensive measurements of various morphological traits led to the emergence of biometrical approaches for quantifying trait heritability, a discipline that was spearheaded by Galton~\cite{Galton_1889}. Soon thereafter, Mendelian genetics began to rise as an alternative approach, with a focus on \textit{qualitative} rather than \textit{quantitative} traits as considered by biometrics. It is only in the early 20'th century that R.A. Fisher provided an elegant reconciliation of this dichotomy: 
in a series of works beginning in 1918~\cite{Fisher_1918,Fisher_1919} and summarized in his book~\cite{Fisher_1930}, he was able to show 
that when \textit{many} genes contribute to a trait, 
Mendelian inheritance leads to a pattern of trait values from one generation
to the next that follows what is seen experimentally and is in agreement with the continuous description given by biometrics. Fisher is thus considered as the founder of the three fields: 
mathematical, population and quantitative genetics.

Fisher formalized this many-genes assumption in a now standard quantitative genetics model
whereby $M$ genes influence the considered trait, and the alleles -- the variants 
associated with a gene -- contribute additively to the trait. His ``infinitesimal model''~\cite{Fisher_1918}
is formally obtained
by taking the limit $M \to \infty$. This limit of an infinite number of genes is particularly
attractive because it allows a thorough mathematical treatment, thanks largely to the fact that
any selection on individuals of an infinite population affects only infinitesimally the allelic frequencies.
The case of \textit{directional selection} is particularly interesting. In such a program to 
improve a population for a particular trait, one first exploits
the variance of the value of that trait in the population,  
selecting the ``best'' individuals. This
selection of course increases the mean of the trait's value in the population
but it also reduces its underlying causal variance, hereafter referred to simply as the \textit{genetic variance}. 
After selection of individuals, the second step of directional selection consists in mating these ``best'' individuals
to produce offspring for the next
generation. This sexual reproduction 
is such that offspring inherit half of their alleles from one parent and the other half from their other parent.
The two steps -- selection followed by sexual reproduction -- form a cycle that is at the heart of
recurrent selection breeding programs. Within each cycle, the genetic variance 
is first decreased because of the selection and it is then increased via the genetic shuffling induced by meiotic crossovers.

The numerous mathematical studies of this system show that
if the population size is infinite and if the inheritance of alleles from parents is random, that is arising independently for the different genes corresponding to 
absence of genetic linkage, then the cycling through generations rapidly converges
to a \textit{steady-state} regime where each generation is improved by a fixed 
amount over the previous
one~\cite{Barton_2017,Bulmer_1971,Bulmer_book,Turelli_2017}. Specifically, the mean value of the
quantitative trait in the population increases at each generation by an amount that is asymptotically
fixed and strictly positive. This behavior is quite appealing because it is coherent with the maintenance of genetic variance
in natural and domesticated populations~\cite{Diamond_2002,Slatkin_1970} and it also justifies the quite linear genetic progress produced in modern directional selection programs for various stocks and crops~\cite{Laurie_2004,Moose_2004}. Moreover the infinitesimal model has found uses for practical situations such as in genomic selection analyses~\cite{Hu_2012,Visscher_2019}.

It is not difficult to see that if one relaxes either of the two standard 
hypotheses of the infinitesimal model (the number of genes $M$ and the 
population size $N$ both being infinite) then the 
asymptotic behavior changes qualitatively with a genetic gain per generation that tends towards 0. Indeed, for finite $M$ (just as for $M=1$), one enters a regime of diminishing returns because the more favorable alleles have a frequency that drops off more and more quickly at each step of selection. Similarly, in the case of finite $N$, all genes ultimately see their alleles become fixed. Thus, using the infinitesimal model to 
justify the substantial genetic variance seen in practice in natural populations 
requires hypothesizing that (i) the population size is sufficiently large 
and (ii) the considered trait depends on sufficiently many genes. These conditions of
``sufficiently large'' and ``sufficiently many'' are relative to the number of generations
over which the infinitesimal model acts: the more generations of directional selection, the more stringently one must meet those two conditions.
It is important to note that this minimal infinitesimal model~\cite{Bulmer_book,Fisher_1918} does not include the appearance of mutations, another mechanism that can maintain genetic variance. 

The infinitesimal model is almost always studied without any genetic linkage because the
mathematical analyses are then much simplified. However, from
a biological point of view, genes are in fact subject to tight linkage for 
the simple reason that they
lie along chromosomes. 
The different studies having tackled the consequences of genetic linkage 
used finite populations~\cite{Keightley_1987,Santiago_1998} or a finite number of genes~\cite{Turelli_Barton_1990,Turelli1994}, in which cases there is no constant long-term gain per generation as we pointed out before. That could justify why all 
 investigations keeping $N$ and $M$ infinite considered that the qualitative behavior of the infinitesimal model is the same whether there is genetic linkage or not: they assumed that in the limit of many generations the system goes to a steady state in which the genetic variance is a strictly positive constant, leading to fixed genetic gain per generation. In effect, according to these studies, linkage simply modifies the steady-state variance and thus the genetic gain per generation.

Our focus in the present work concerns whether that default 
assumption
\cite{Gallais_2003,Walsh_Lynch_2018}
is correct. Interestingly, in spite of the fact that Fisher's infinitesimal model 
was formulated over a century ago, there has been no mathematical progress in treating the 
$M = N = \infty$ limit. Given that in biological settings one has many tightly linked genes lying along chromosomes, in this work we connect to the infinitesimal model by taking the continuous limit. Specifically, we distribute an infinite number of genes along one continuous chromosome (for simplicity). The natural mathematical framework for treating that system relies on Fourier series which we then analyze based on the standard Gaussian approximation. As a result, we
find that in contrast to expectations,
the genetic variance decreases to 0 as the number of generations increases. There is thus no analog
of the steady-state behavior arising in the absence of genetic linkage. Instead,
a more singular behavior arises whereby the genetic variance goes to zero
following a subtle scaling law characteristic of systems undergoing very slow or ``aging'' dynamics~\cite{Barrat_Mezard_1996,Bouchaud_1996,Kaplan_Balaban_2021,Scherer_1986,Struick_1978}.

\section{The infinitesimal model without linkage}
\subsection{Hypotheses}
Fisher's infinitesimal model begins with a quantitative trait controlled by a large number $M$ of genes. Each individual (plant or animal) has a genotype embodied by the allelic content of those genes. The cycle of sexual reproduction alternates between a haploid and a diploid phase. In the haploid phase, one has \textit{gametes} with a single copy of each chromosome. Each gamete's genotype is specified by a list, with one allele for each gene. In its simplest version, the genetic value is defined at the gametic level and is taken to be
the sum of the values contributed by each allele:
\begin{equation}
G = \sum_{m=1}^{M} x_m
\label{Eq.G}
\end{equation}
where $x_m$ is the (additive) 
effect of the 
allele carried by gene $m$ ($1 \le m \le M$) in the considered gamete. When analyzing the passage from one generation to the next, one can focus on the haploid or the diploid phase of the cycles depending on when selection arises. In the present work we will follow the content of the haploid phase as it allows the compact representation of Eq.\ref{Eq.G} and thus a much simpler mathematical treatment. However, this comes at the cost of allowing selection only in the haploid phase rather than in the diploid phase or even in both phases. As a consequence, we cannot treat a number of effects such as genetic dominance.

The values of $G$ are partly heritable from one cycle to the next
because of the allelic shuffling produced during meiosis (taking one from the diploid to the haploid phase). Across generations, the allelic content in the population of gametes changes because of selection, but the values of the alleles themselves are fixed. Those values are specified in the gametes of the first generation. A simple choice is to assume alleles have just two possible values, for instance $\pm 1/\sqrt{M}$. Another common choice is to allow an arbitrary number of alleles with values drawn independently from a Gaussian distribution of zero
mean and variance of $1/M$. (These dependencies on $M$ ensure that the initial variance of $G$ is 1 for all values of $M$.) As $M \to \infty$, the two choices are equivalent because of the central limit theorem. For convenience, we will use the second choice for the rest of our analysis. 

At the first generation, $G$ is a Gaussian random variable of zero mean and unit variance, but over successive generations its mean rises because of the selection; for the present work, we 
assume
the selection to arise during the gametic phase, operating via the value of $G$. Thus each generation is produced from the previous one according to the cycle: (i) one selects the gametes (haploid genotypes) having high $G$ values in the current population and (ii) one produces the next generation by panmictic mating, \textit{i.e.}, by choosing pairs of gametes at random among the selected ones and ``recombining'' them into a single new gamete via allelic shuffling. The hypothesis of \textit{no genetic linkage} corresponds to having independent assortment of the alleles when going from a pair of gametes to the next generation gamete. Given these choices, the infinitesimal model is formally obtained by taking the limit whereby 
both the number of genes $M$ and the population size $N$ go to $\infty$.

\subsection{The steady-state property}
In most breeding programs, the selection phase typically keeps only those individuals whose (apparent) genetic value is above a generation-dependent threshold. Let $\mu$ and $\sigma$ be the mean and standard deviation of $G$
within the current population; common practice is to then 
select those gametes whose $G$ value is greater than $\mu + \beta \sigma$ where 
$\beta$ is a measure of the 
selection intensity. At large $\beta$ only a small fraction of the population of gametes is used for mating, and in practice the distribution of $G$ falls off rapidly so that most selected gametes have their $G$ value very close to the imposed threshold. A major mathematical simplification arises in the infinitesimal model if one selects instead the 
gametes whose $G$ value is \textit{exactly} at that threshold (this is possible because the population size is infinite). With those choices, let us now review~\cite{Barton_2017,Bulmer_book,Turelli_Barton_1990,Turelli1994} the process of
going from the initial population to the next generation, for fixed $M$ but then taking $M$ to infinity. 

In the initial population (generation $0$), all $x_m$ are \textit{i.i.d.} 
Gaussian random variables of mean $0$ and variance $1/M$, so the joint
distribution of the $x_m$ is multi-Gaussian and the associated covariance matrix is diagonal.
After selection, since we impose $G=\mu_0 + \beta \sigma_0$
($\mu_0=0$ and $\sigma_0=1$ in the initial generation),
the joint distribution of the $x_m$ is modified but it remains multi-Gaussian because the selection corresponds to imposing a \textit{linear} constraint on the variables. What are the parameters of this new (multi-Gaussian) distribution? Clearly, by using the permutation symmetry amongst the genes, all the means must be increased by $\beta \sigma_0 / M$. More subtly, the selection introduces a dependency amongst the $x_m$ (since their sum is fixed) and thus the off-diagonal covariances no longer vanish: selection has introduced ``linkage disequilibrium",
a phenomenon
that in this context it is called
the Bulmer effect (\textit{cf.}~\cite{Bulmer_1971}). By direct calculation~\cite{Barton_2017,Bulmer_book,Turelli_Barton_1990}, e.g., using the discrete Fourier transform of the $x_m$, one finds that the off-diagonal covariances are all equal to $-1/M^2$ while the variances are all $1-1/M$. Of course these values ensure that the variance of $G$ after selection is $0$.

Moving on to the sexual reproduction phase of the cycle, one takes pairs of gametes and performs random assortment, \textit{i.e.}, the first gamete transmits its alleles for a subset of $M/2$ randomly chosen genes while the second one transmits its alleles for the other subset of $M/2$ genes. When going from all $x_m$ to a subset thereof, the joint distribution of the $x_m$ in this subset is again multi-Gaussian. Furthermore, the hard constraint on the sum of the $x_m$ is lost when one considers a subset containing only half of those $x_m$. In fact, as can be seen by direct calculation, the sum of the kept $x_m$ in each case is a Gaussian variable of variance $1/4$ (\textit{cf.}~\cite{Barton_2017}). Since the two gametes are independently drawn from the infinite population, the $G$ value for the ``recombined'' gamete also has a Gaussian distribution, and its new variance is $1/2$ (to be compared with the value 0 after selection). This ``boosting'' of the variance of $G$ by the assortment phase is referred to as the \textit{release} of genetic variance; in effect, the random assortment is able to mine the (infinite) variability available in the (infinite number of) $x_m$. 

To summarize this $M \to \infty$ limit, the distribution of the $x_m$ in the population of recombined gametes goes to a multi-Gaussian where the $x_m$ can be thought of as (quasi) independent variables of mean $\beta \sigma_0/M$ and variance $1/M$ but subject to the constraint that $G$ is of variance $1/2$ ~\cite{Barton_2017,Bulmer_1971,Bulmer_book,Lange_1978}. Interestingly, with our choice of selection, we reach the steady state of the cycling in just one generation. Specifically, at generation $k\ge1$, the joint distribution of the $x_m$ is a multi-Gaussian such that the $x_m$ can be considered to be independent Gaussian random variables of variance $1/M$, but subject to the constraint (hidden in covariances that are of order $1/M^2$ and are thus naively ``lost'' when $M \to \infty$) that the variance of their sum (\textit{i.e.}, the variance of $G$) is $1/2$. At each generation, the selection phase of the cycle makes the variance of $G$ go to 0 but then the phase with random assortment restores a variance of $1/2$. One thus has a steady state behavior when comparing one cycle to the next, all statistical properties are constant except for the means: those of the $x_m$ increase by $\beta /M\sqrt{2}$, and that of $G$ increases by $\beta /\sqrt{2}$.

\section{The infinitesimal model \textit{\normalsize{with}} linkage}
\subsection{General framework and model choices}
Genetic linkage in real biological situations arises because genes belong to chromosomes so the transmission of an allele at one position of a chromosome generally leads to the simultaneous transmission of a large surrounding region, producing strong correlations of inheritance along chromosomes. Numerous studies have considered a finite number of linked loci~\cite{Buerger_book,Felsenstein_1965,Hill_1966,Qureshi_1968,Robertson_1970,Turelli_Barton_1990,Turelli1994,Walsh_Lynch_2018} showing that linkage generally reduces the response to selection, but so far a controlled treatment directly within the infinitesimal model (having an infinite number of linked loci) has not been provided.

\begin{figure}[h]
\centerline{\includegraphics[width=1\textwidth]{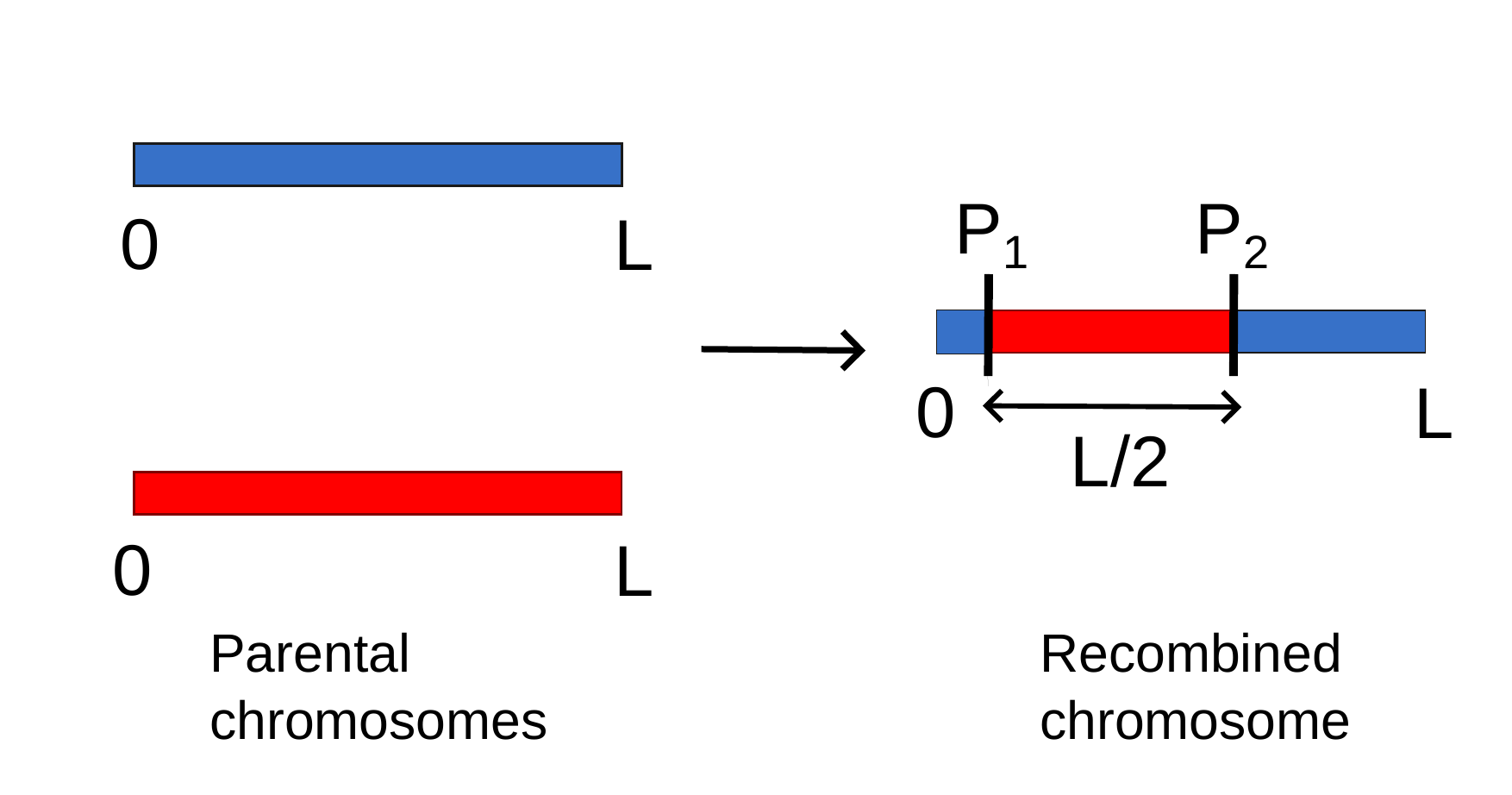}}
\caption{Assortment of alleles with linkage: in our framework, each gamete consists of one continuous chromosome 
forming a segment of length $L$. During the assortment phase, two 
parental gametes (in blue and red) ``recombine'' so the new 
(child) gamete inherits half of its genetic content from each gamete of that pair. Crossover points are labeled $P_1$ and $P_2$. 
}
\label{Fig.crossovers}
\end{figure}

To simplify our analysis of the effects of genetic linkage, we shall consider that the gametes carry just one chromosome. Furthermore, to facilitate the mathematical derivations, we will take that chromosome to be continuous~\cite{Slatkin_1972}. To go from the current generation to the next, each ``recombined'' chromosome of the next generation is taken to be an assortment of two chromosomes of the current generation. As illustrated in Fig.\ref{Fig.crossovers}, we implement this combination by introducing two crossovers, say at positions $P_1$ and $P_2$. The recombined chromosome has its $[P_1,P_2]$ region coming from one of those two current-generation chromosomes. If we interpret positions beyond the chromosome's end using a periodic representation, we can consider that the
 $[P_2,P_1]$ region 
is also an interval
coming from the other
of the two current-generation chromosomes. For our modeling of meiosis, we
 impose those two regions to be of the same
total size. With the use of the interpretation extending the chromosome periodically, in effect one can consider
 $P_1$ to be uniformly distributed along the chromosome, 
corresponding to a recombination rate that is constant, independent of position.

\subsection{Introducing the Fourier series representation}
We continue to assume that the population size is infinite. The position $s$ ($0 \le s < L$) 
of a gene 
on the chromosome can be conveniently replaced by the quantity 
$\theta=2 \pi s/L$ ($0 \le \theta < 2\pi$) which can then be interpreted as an angle.
One can specify the genotype of a gamete either by its 
allelic values 
$x(\theta)$
along the 
chromosome
or by the corresponding representation in Fourier space. In the limit $M\to\infty$, this representation takes one from the profile $x(\theta)$ to the Fourier components labeled by $k$:
\begin{equation}
X_k = \frac{1}{2\pi} \int_{0}^{2\pi} e^{-ik \theta}x(\theta)d\theta
\end{equation}
where $k$ runs over all integers, from $-\infty$ to $+\infty$. Since $X_{-k}$ is the complex conjugate of $X_k$, which we denote $\overline{X_k}$, any genotype can be specified by $\mathbf{X}=\{X_k, 0 \le k < +\infty\}$. Note that $X_{k=0}$ is real. 
If one applies the inverse transform, one recovers the profile (which is periodic in $\theta$):
\begin{equation}
x(\theta) = \sum_{k=-\infty}^{+\infty} e^{i k \theta} X_k
\label{Eq.fourier_series}
\end{equation}

In
the initial population
satisfying linkage equilibrium,
the $x(\theta)$ are \textit{i.i.d.} random variables of zero mean, corresponding to Gaussian white noise so that $E[x(\theta) x(\theta')]$ is the Dirac delta function $\delta(\theta-\theta')$. Transforming to Fourier space, we see that the real and imaginary components of $\mathbf{X}$ are independent, all means are $0$ and all variances are $1/2$ except that of $X_{k=0}$ which is $1$. The vector $\mathbf{X}$ specifying a genotype in this initial population thus follows a multi-dimensional complex normal (Gaussian) distribution \cite{Goodman_1963,Lapidoth_2017,Picinbono_1996}. All means vanish and the covariance matrix is the diagonal identity matrix:
\begin{equation}
E[(\overline{X_k - E[X_k]})(X_{k'}-E[X_{k'}])] = \delta_{k,k'}
\label{Eq.complexNormal}
\end{equation}
where $\delta_{k,k'}$ is the Kronecker delta ($k \ge 0$ and $k' \ge 0$). The ``pseudo-covariance'' matrix (like Eq.\ref{Eq.complexNormal} but without the complex conjugate) is zero everywhere except for the (0,0) entry which is 1.

\subsection{
Selection and recombination in Fourier space
}
Let us now
consider how the $X_k$ are changed as one performs one cycle of directional selection, applying first the selection of gametes as was done in the case without linkage and then recombining them in pairs as shown in Fig.\ref{Fig.crossovers} to produce recombined gametes of the next generation. 
Not surprisingly, the
selection of gametes having a given genetic value $G$ simply corresponds to collapsing the distribution of $X_{k=0}$ to a Dirac delta function at that value, leaving all other aspects of the distribution unchanged. (This is the analog of what we reviewed for the case without linkage.) 
In the second part of the cycle, there is panmictic mating of the selected gametes. For a given pair of gametes of genetic content $\mathbf{X}$ (the transform of $x(\theta)$) and $\mathbf{X'}$ (the transform of $x'(\theta)$), define $\psi$ to be the angle ($0 \le \psi < 2 \pi$) of the point $P_1$ in Fig.\ref{Fig.crossovers}. The allelic profile of the recombined gamete is the sum of two contributions, $x(\theta)H(\theta-\psi)$ and $x'(\theta)H(\theta-\psi-\pi)$, where $H$ is the periodic step (Heaviside) function that has the value 1 in all intervals $[2n\pi,(2n+1)\pi]$ and vanishes everywhere else ($n$ can be any integer). The key point is that these terms are linear functions of $\mathbf{X}$ or $\mathbf{X'}$. (In Appendix I we provide the explicit expressions for the recombined gamete's Fourier coefficients as a function of the ones in the recombining pair.) Noting that linear combinations of multi-dimensional complex normal variables are themselves multi-dimensional complex normal, we can conclude
by recursion
that at each generation 
each genotype follows
a multi-dimensional complex normal distribution, with a covariance matrix that is a function of the successive crossover positions leading to that gamete across generations. 
Consequently, 
conditional on those crossover positions,
the joint distribution of the $X_k$ is 
multi-complex normal
at all generations.

\subsection{The complex normal projection and inter-generational recursion relations}
Since the crossover positions are random, the joint distribution of the $X_k$ over the \textit{whole} population is a mixture of multi-complex normal distributions. To progress, we project that mixed distribution back to a multi-complex normal distribution, for which we need to compute the covariance and pseudo-covariance matrices.
Although it may seem challenging to 
do so,
a very useful feature follows from the invariance of the problem under rotation, \textit{i.e.}, $\theta \to \theta + \delta$ for any shift $\delta$. Indeed, the initial population is invariant under rotations and this invariance is preserved from one generation to the next. That constrains enormously the mathematical equations because when performing a rotation by an angle $\delta$ of the profiles of all gametes in the population, each $X_k$ will be multiplied by a phase $e^{-ik\delta}$ showing that the mean of $X_k$ vanishes if $k\ne 0$. Similarly, the covariance of $X_k$ and $X_{k'}$ will be multiplied by $e^{-i(k'-k)\delta}$, showing that the covariance matrix must be diagonal. Applying the same reasoning to the pseudo-covariance matrix, we see that it has to vanish except for its (0,0) entry. Thus,
within this projection approximation,
to follow the multidimensional distribution of $\mathbf{X}$ when going from one generation to the next, we only need to determine how the mean of $X_{k=0}$ changes and how the (complex) variances of the $X_k$ are modified. We do this by direct calculation from the expression for a recombined gamete's Fourier coefficients in terms of that of the two contributing gametes (that derivation is given in the Appendix I) followed by the calculation of the associated variances (that derivation is given in Appendix II). We summarize the results as follows.

Denote by $\mu_{g}$ the mean of $X_0$ at generation $g$. Similarly, denote by 
$\sigma^{2}_{k, g}$ the (complex) variance of the $k$-th 
Fourier coefficient at generation $g$, that is the expectation of $\overline{X_k} X_k$ (which is real). One then has:
\begin{equation}
\mu_{g+1} = \mu_{g} + \beta \sigma_{0, g}
\label{Eq.mean}
\end{equation}
in direct analogy with what occurs in the absence of linkage, $\beta$ still being the parameter associated with the selection 
intensity.
Note that the selection for a given value of $G$ affects 
only the Fourier component at $k=0$, setting it to zero.
Then, including the effects of the second phase of the cycle during which alleles are shuffled by
crossovers as shown in Fig.\ref{Fig.crossovers}, we obtain the following recursions for the variances which are most easily written using both positive and negative indices for the Fourier coefficients ($-\infty < k, k' < \infty$):
\begin{equation}
\sigma^{2}_{k, g+1} = \dfrac{\hat{\sigma}^{2}_{k, g}}{2} + \dfrac{2}{\pi^{2}} \sum_{k' odd} \dfrac{\hat{\sigma}^{2}_{k+k', g}}{k'^{2}} 
\label{Eq.recursion}
\end{equation}
where $\hat{\sigma}_{k,g} = \sigma_{k,g}$ if $k \ne 0$ while 
$\hat{\sigma}_{0,g} = 0$ because the variance of $G$ is set to 0 during the selection phase.
In Eq.\ref{Eq.recursion} the sum over $k'$ is to be taken over both positive and negative odd integers (note however that $\sigma_{k,g} = \sigma_{-k,g}$). The initial conditions for these recursions are $\sigma_{k, g=0}^2 = 1$ for all $k$, corresponding to 
linkage equilibrium
Furthermore,
one also has the boundary condition for all $g$: $\sigma_{k,g}^2 \to 1$ as $k\to \pm \infty$. 
Eqs.\ref{Eq.mean} and \ref{Eq.recursion} then 
specify within the complex normal projection approximation
the change in the distribution of $\mathbf{X}$ values in the population when going from one generation to the next, in the limit of an infinite number of genes and an infinite population size. 
Even after a single iteration of Eq.\ref{Eq.recursion}, the variances will no longer all be equal, corresponding to the appearance of linkage disequilibrium that is expected to build up over generations.
Nevertheless, in
the standard lore, one expects $\sigma_{0, g}^2$ (often referred to as the additive genetic variance) to go to a limiting strictly positive value at large $g$. 


\subsection{Absence of a steady-state solution
to the recursion equations
}
The recursions for the variances in Eq.\ref{Eq.recursion} have the following interpretation if one thinks of $\sigma_{k,g}^2$ as specifying the amount of ``matter'' at site $k$ and generation $g$. When going from $g$ to $g+1$, half of the matter at each site stays put while the other half is shared across other sites. That sharing is done via a ``diffusion'', implemented mathematically via a convolution kernel. In effect, a fraction $4/(\pi^2 k'^2)$ of that transferred matter is assigned to a site at a distance $|k'|$, where $k'$ (positive or negative) is imposed to be odd. (Of course the sum of all those fractions is 1.) Note that this diffusion process conserves the total amount of matter. However the case $k=0$ includes a resetting whereby all matter at that site is first removed, thereby leading to loss of matter. Therefore, the amount of matter lost at each iteration -- at the level of the whole system -- is precisely the value of $\sigma_{k=0,g}^2$. Each iteration is thus an alternation of (i) removing all the matter at site $k=0$ and (ii) diffusing matter across all sites with the aforementioned convolution kernel. In analogy with the case without linkage, we may expect that there is a steady-state solution to these recursions to which the system converges as $g \to \infty$.

Before treating the true convolution kernel arising here (it is spread out to infinity, thereby complicating the analysis), consider first the simpler situation where one replaces it by the standard diffusion convolution kernel. In that case, the transferred matter is equally shared between one's nearest neighbors on each side (so $k'$ takes on only the values $\pm 1$). Assume that there exists a steady-state for this system, \textit{i.e.}, that there is a set of $\sigma_{k,*}^2$ which are invariant under the processes of (i) removal of all matter at $k=0$ and (ii) sharing half of one's matter between one's nearest neighbors. The steady state condition then imposes that the matter transferred ``out'' 
of any given site
is exactly compensated by an equal amount of matter transferred ``in''. For any $k \ne 0$, that requires $\sigma_{k,*}^2$ to be equal to the average of $\sigma_{k-1,*}^2$ and $\sigma_{k+1,*}^2$. As a consequence, on each side of the origin ($k=0$), $\sigma_{k,*}^2$ must be linear in $k$. However, that property necessarily contradicts the boundary condition $\sigma_{k,*}^2 \to 1$ as $g \to \infty$, from which we conclude that the initial hypothesis (existence of a steady state) must be false. Instead, the system will exhibit non stationary dynamics, whereby at any finite $g$ the $\sigma_{k,g}^2$ will smoothly interpolate on each side of the site $k=0$ between a near vanishing value and the value $1$ far away, and as $g$ increases, the overall region of very low values will become wider and wider, corresponding to an unbounded build-up of linkage disequilibrium.

The impossibility of a steady-state behavior can also be shown in the continuum scaling limit using a Green function approach. To do so it is convenient to change variables from $\sigma_{k,g}^2$ to $c_{k,g} = 1-\sigma_{k,g}^2$ and again interpret the variables as amounts of matter. The corresponding boundary conditions are now $c_{k,g} \to 0$ as $k\to \pm \infty$ for all $g$. Furthermore, the step (i) becomes (i') ``set $c_{k=0,g}$ to 1'' while the step (ii) remains unchanged. The advantage of this change of variables is that the total matter present in the system is finite and so can be more easily followed. In particular, it starts at 0 (\textit{cf.} the initial conditions) and it can only increase at each step. A cycle (one generation) then corresponds to first resetting the value of $c_{k=0,g}$ to 1 and then applying the diffusion step whereby for each site half of its matter is transferred to other sites. Suppose we apply (i') and (ii) starting from the initial conditions and then follow the behavior of the $c_{k,g}$ at large $g$. Given the definition of the $c_{k,g}$, they will always be less or equal to 1. At each step, some amount of matter is introduced at the site $k=0$ and then diffusion is applied. Each amount of matter added at generation $g_1$ at the site $k=0$ will lead to a spread-out distribution at a later generation $g_2$, and in the continuum limit when $g_2-g_1$ is large that distribution corresponds to the Green function which has a Gaussian form: $\sqrt{\pi (g_2-g_1)} \exp[-k^2/(g_2-g_1)]$ (the ``missing'' factor of 2 in the diffusion constant follows from the fact that at each generation only half of the matter at each site participates in the diffusion). Using the linearity of the equations, this Green function approach then shows that the total response is obtained by summing this last expression over all $g_1$ values. If the system converges to a putative steady state where $c_{k=0,*}<1$, then the added amount of matter at each step tends towards a fixed and strictly positive amount, from which we obtain that $c_{k,g_2} \sim \sqrt{g_2}$ at large $g_2$. That contradicts the fact that $0 \le c_{k,g_2} \le 1$. Thus if the system converges to a steady state distribution, it must satisfy $c_{k=0,*}=1$. But if that is the case, the steady state condition on the distribution implies $c_{k,*}=1$ for all $k$, which does not respect the boundary condition $c_{k,*}\to 0$ as $k \to \pm \infty$. This Green function approach thus shows, just like the previous approach, that the system cannot converge to a steady state at large $g$.

This last method of analysis allows us to now treat the original recursions Eq.\ref{Eq.recursion} for which the diffusion convolution kernel decays as $1/k'^2$. This convolution kernel has the property that its moments (expectation of $k'^\alpha$) are finite if and only if $\alpha < 1$. This exponent $\alpha$ is characteristic of the so-called $\alpha$-stable distributions\cite{stable_distributions_1981} to which our convolution kernel will converge to in the continuum limit. Specifically, based on the generalized central limit theorem~\cite{Samorodnitsky_book},
the iteration of this convolution kernel at large $g$ leads in the continuum limit to the Cauchy distribution $\pi^{-1} (2g/\pi)/((2g/\pi)^2 + x^2)$. This is the direct analog of the Gaussian distribution of the previous paragraph and it provides the Green function of the problem. We thus repeat the previous reasoning using again the linearity of the recursions. If we assume that the iterations converge to a steady state behavior for which $c_{k=0,*}<1$, by summing over all contributions produced by the added matter at $1 \le g_1 \le g_2$ we obtain $c_{k,g_2} \sim \ln{(g_2)}$ at large $g_2$. This diverges when $g_2 \to \infty$, contradicting the fact that $0 \le c_{k,g_2} \le 1$. Thus in the steady state one must have $c_{k=0,*}=1$, but as before that leads to $c_{k,*}=1$ for all $k$ which is not acceptable given the boundary condition $c_{k,*}\to 0$ as $k \to \pm \infty$. So we can conclude that the recursions Eq.\ref{Eq.recursion} lead to a non-stationary behavior, and that as $g$ increases, $\sigma_{k,g}^2$ converges to 0. We now turn to the numerical analysis of Eq.\ref{Eq.recursion} to study \textit{how} $\sigma_{k=0,g}^2$ tends to 0 with increasing $g$.


\subsection{Numerical treatment
of the recursion equations
}
To treat numerically the infinite set of recursions given in Eq.\ref{Eq.recursion}, we have to focus on a finite number of variances and in practice we consider those having index $k$ with $|k| \le N$, $N$ large. 
Each recursion involves all $k'$ and so, in our numerical treatment, the variances beyond the cut-off $N$ are replaced by analytic expressions that are set by matching the values in the neighborhood of $k = \pm N$. This avoids the severe truncation error that would have followed if we had simply replaced those variances at $|k| > N$ by their limiting value. Naturally, the results need to be extrapolated to the $N=\infty$ limit as best as possible. In practice, having a finite $N$ introduces errors that become significant when $g$ has the same order of magnitude as $N$. Thus it is necessary to perform a study of the numerical behavior of $\sigma^{2}_{0, g}$ as a function of $N$.
Noticing that the recursions involve a convolution, we compute the convolution using the discrete Fast Fourier transform. This allows us to treat $N$ values up to tens of thousands on a desktop computer. We now discuss the results obtained with these numerical methods. The computer codes (R \cite{Rsoft} scripts for implementing efficiently the numerical iterations of Eq.\ref{Eq.recursion} and for the corresponding analyses) are provided as Supplementary Material.


\subsection{The system displays aging dynamics as the generation number increases}
As mentioned before, the initial condition is $\sigma_{k,g=0}^2=1$, and one also has the boundary condition $\sigma_{k,g}^2=1$ when $k \to \infty$ for any $g$. For numerical purposes, we change variables and follow instead $1-\sigma_{k,g}^2$ as a function of $g$. That is more convenient because then these quantities go to 0 as $|k| \to \infty$, at any fixed $g$, and so problems of numerical round off arise for much larger $N$ and $g$ than without that change.

\begin{figure}[h]
\centerline{\includegraphics[width=1\textwidth]{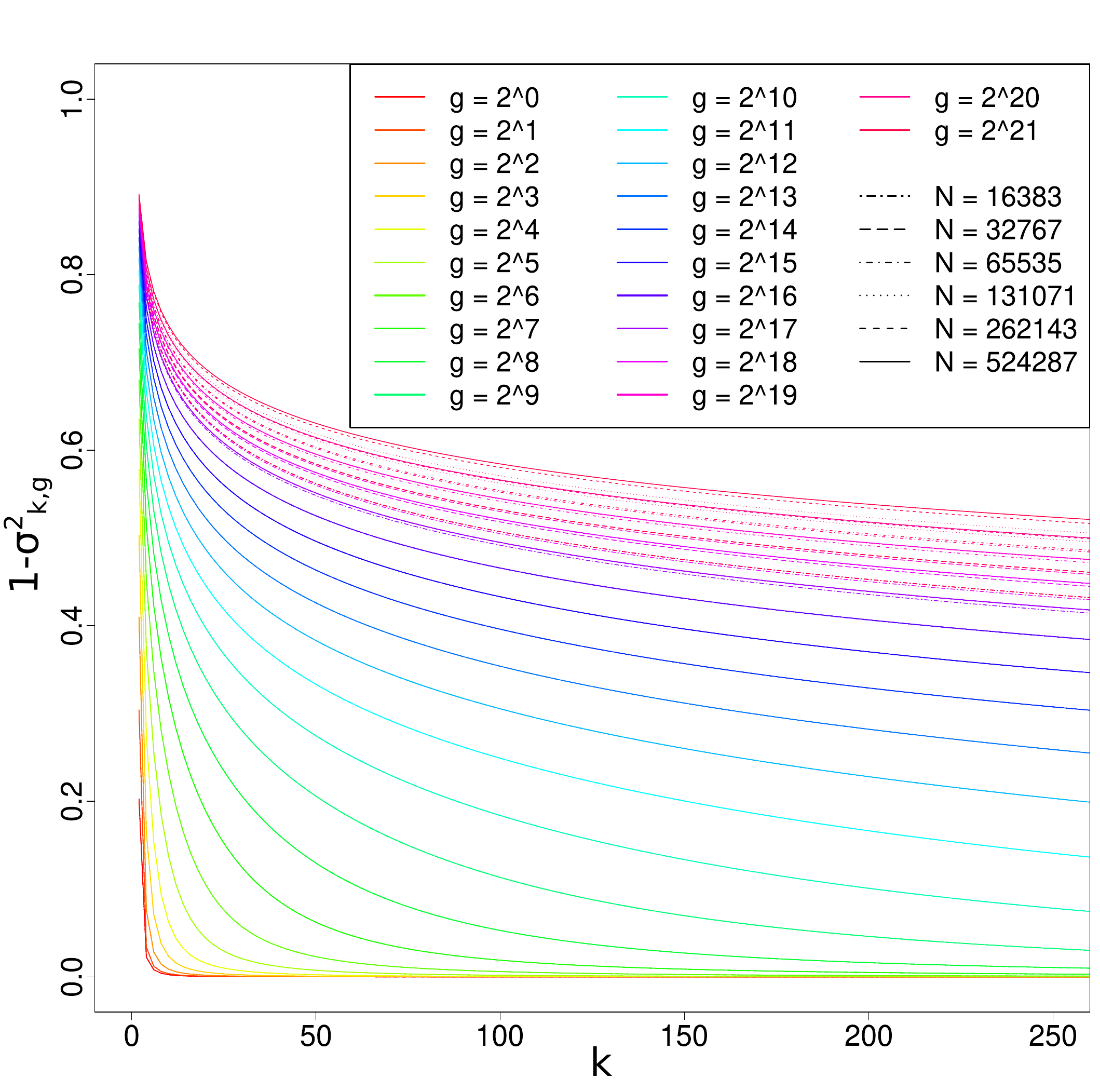}} 
\caption{Evolution of the first 250 variables ($1-\sigma_{k,g}^2$ for $0\le k \le 250$) at increasing $g$ values. Note that the errors introduced by the truncation parameter $N$ become visible when $g$ is comparable to $N$.}
\label{Fig.variances}
\end{figure}

Fig.\ref{Fig.variances} displays our results for the first 250 variables, specifically $1- \text{variance}$ for $0 \le k \le 250$, taking a number of values of $g$. Clearly all these quantities decrease with increasing $g$, but this occurs very slowly in a way typical of aging dynamics~\cite{Barrat_Mezard_1996,Bouchaud_1996,Kaplan_Balaban_2021} as evidenced by the fact that values in this figure are displayed for $g$ values equal to powers of $2$. In the context of the explanation given in the ``Absence of a steady-state solution'' section, such an aging behavior is not a surprise. Indeed, the matter added at each generation at the $k=0$ site decreases to 0 with generations while keeping the system away from a steady state.

We also see that the errors introduced by the truncation parameter $N$ become visible only when $g$ is comparable to $N$. These slow dynamics would be quite difficult to identify in a forward simulation of a population subject to directional selection. In particular, to ensure small enough drift effects, it would be necessary to simulate huge population sizes.

\subsection{The genetic variance decays to 0 inversely with the logarithm of generation number}

\begin{figure}
\centerline{\includegraphics[width=1\textwidth]{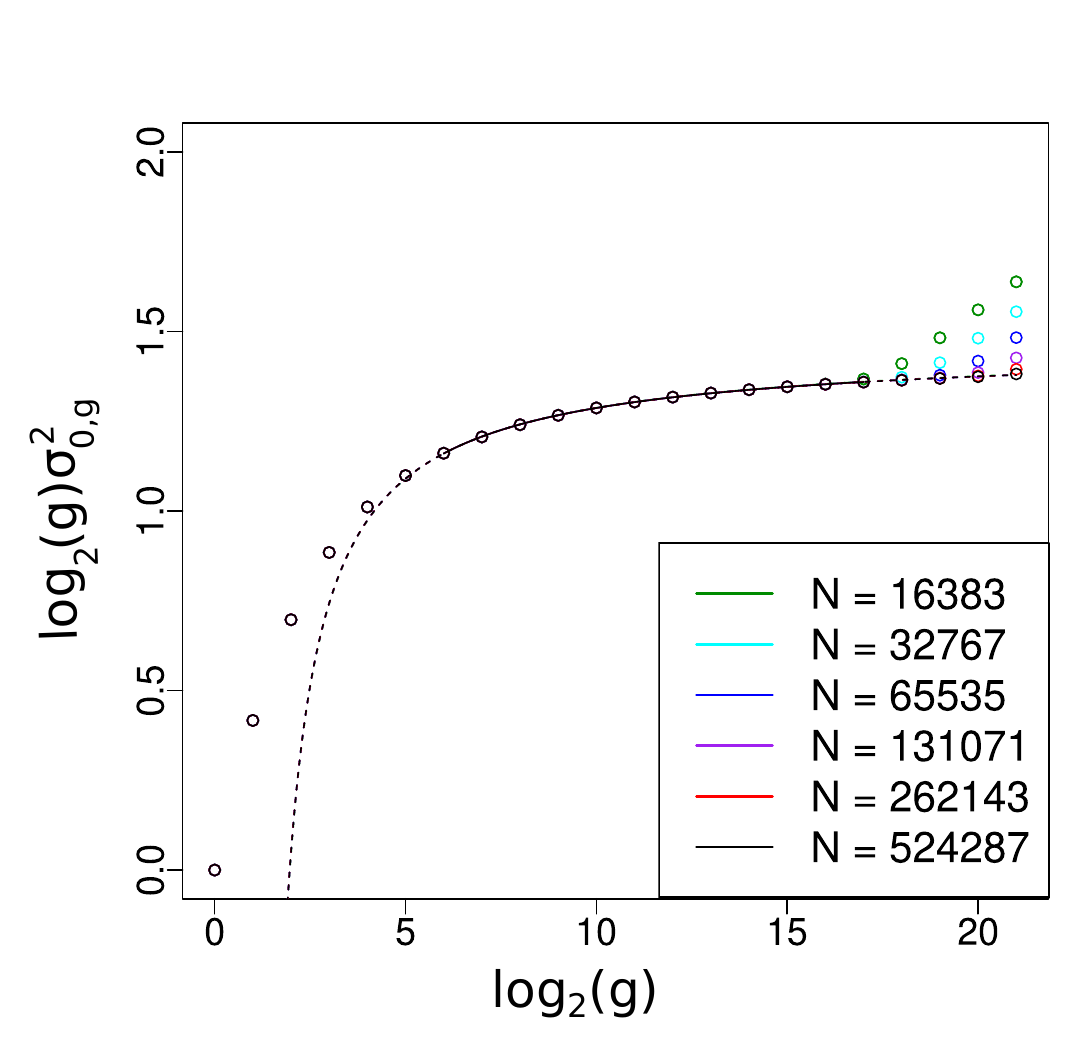}} 
\caption{Scaling law and corrections for the genetic variance $\sigma_{k=0,g}^2$ when the number of generations $g$ becomes large. Y axis is $log_2(g) \sigma_{0,g}^2$ estimated using different values of $N$. The numerical data at finite $N$ is shown via dots and our associated fit by the continuous curve (regions not used for the fit are indicated by a dotted line). X axis is $log_2(g)$.
The function for the fit uses two parameters $A$ and $B$ and is specified in Eq.\ref{Eq.scaling}.}
\label{Fig.scaling}
\end{figure}

The $k=0$ variance component determines the genetic gain (cf. Eq.\ref{Eq.mean}) which is the main observable of interest in quantitative genetics. Our scaling analysis of $\sigma_{0,g}^2$ is summarized in Fig.\ref{Fig.scaling} where we display the product $log_2(g) \sigma_{0,g}^2$ as a function of $log_2(g)$. The curves for different values of $N$ superpose well except that finite $N$ effects are visible at large $g$, in direct analogy with what was found in Fig.\ref{Fig.variances}. As $N$ increases to infinity, we see empirically that the curves tend to an envelope that is well fit to the function
\begin{equation}
log_2(g) \sigma_{0,g}^2 = A + \dfrac{B}{log_2(log_2(g))\sqrt{log_2(g)}}
\label{Eq.scaling}
\end{equation} 
with $A=1.48$ and $B=-2.02$. From this we conclude that the genetic variance decreases to 0 as the inverse of the logarithm of $g$,  with higher order corrections that include a logarithmic correction factor (thus a log log term) reminiscent of correction to scaling terms arising in various one-dimensional moving front problems~\cite{Berestycki_2018}.

%

\subsection{The long-term genetic gain is unbounded}
Since the genetic variance goes to 0 at large $g$, Eq.\ref{Eq.mean}
shows that the gain per generation goes to zero. Nevertheless, 
the total gain, accumulated over generations, goes to infinity 
with increasing $g$. Indeed, by 
Eq.\ref{Eq.mean}, that total gain is
$\beta$
times the sum over $g$ of the $\sigma_{0,g}^2$. The leading term in Eq.\ref{Eq.scaling} shows 
that $\sigma_{0,g}^2 \approx A/log_2(g)$, and so it is easy to see that the corresponding
series is divergent. A qualitative way to justify this property
is to notice that at any $g$ there remains genetic variance in
the $\sigma_{k,g}$ at large $k$ as in the case without linkage and that its variance can be transferred to $k=0$ over generations. It is just a question of applying sufficiently many crossovers (generations) to recover that reservoir of variance. As a result, the reachable total genetic gains are the same whether there
is linkage or not, it is just that with linkage it takes far more generations to access that reservoir and in fact one has a case of ``diminishing returns'' per generation: it takes more and more generations to provide a fixed genetic gain.

\section{Discussion and conclusions}
We have revisited the infinitesimal model, originally proposed by Fisher in 1918~\cite{Fisher_1918}, with the goal of determining the consequences of linkage therein.
Indeed, in spite of decades of work on that model, the challenge of quantitatively treating linkage amongst an infinite number of genes has never been squarely addressed. To do so, we
first introduced a number of model choices which allowed us to derive exact recursion equations 
for the allelic Fourier coefficients from one generation to the next (cf. Appendix I). Then, using the projection onto complex-normal distributions, we computed the recursions for the means and the covariance matrices (Eqs.\ref{Eq.mean} and \ref{Eq.recursion}). 
The analysis of
these equations
revealed a qualitatively different long-term behavior compared to the case without linkage: the genetic variance (the variance of $G$ in Eq.\ref{Eq.G}, equal to $\sigma_{k=0,g}^2$ in our notation) goes to 0 at large $g$ and so does the genetic gain per generation (cf. Eq.\ref{Eq.mean}). The subtlety of this process likely explains why no previous work had suggested that introducing linkage might qualitatively change the behavior of the infinitesimal model. Furthermore,  our numerical treatment shows that the corresponding dynamics exhibit aging; as a result, any simulation based on following populations of individuals will almost certainly be inconclusive. In effect, although the constraint of linkage is moderate at each generation, when considering many generations, its cumulative effect
increases indefinitely the linkage disequilibrium and thereby changes the long-term behavior of the system in a singular way. 

At the mathematical level, the origin of this singular behavior lies in the way the diffusion convolution kernel decays with distance. For an inverse power law of exponent 2 (as arises in the present case) and also for faster rates of decay (e.g., as occurs for the standard nearest-neighbor diffusion convolution kernel that was considered for pedagogical purposes in the ``Absence of a steady-state solution'' section), the reduction of variance produced in the low frequency modes does not diffuse away sufficiently well into the high frequency modes to prevent the ultimate vanishing of all variances at large $g$. Were the rate of decay \textit{slower}, say according to an inverse power $1+\alpha$ ($0<\alpha<1$), a non trivial steady state might emerge instead, but realizing that type of decay using crossovers would require both an infinite number of crossovers and introducing strong statistical dependencies between them. Interestingly, the case without linkage can be thought of as an extreme limit of such a rate of decay: the infinite number of crossovers required to break all linkage corresponds to a convolution kernel that diffuses out to infinity in an instantaneous manner, so the variance reduction produced at the level of the genetic variance is diluted across all modes in a single step, and thus there is no finite effect on any individual mode.

Our mathematical treatment took advantage of having an infinite population and an infinite number of genes as assumed by Fisher. But it also relied on several choices staying within the infinitesimal model: applying the selection step in the haploid phase
(as a result we can't treat effects of dominance),
having a single 
chromosome per gamete, imposing exactly two crossovers that produce equal sized segments in the genetic recombination step
thereby forcing a flat recombination landscape and a fixed genetic length,
and finally selecting gametes of a given genetic value $G$. Treating other model choices, such as applying selection during the diploid phase as usually done~\cite{Otto_2018} and allowing multiple linear chromosomes, would certainly introduce major technical difficulties for both the mathematical and numerical analyses. Nevertheless, it should be clear that the key property 
driving the unlimited build-up of linkage disequilibrium
is the rate at which the variance reduction (originating at the level of $G$ during the selection step) diffuses out towards the high frequency modes (\textit{i.e.}, the small scales at the level of the chromosomes). The
exponent of the
power law in the corresponding diffusion convolution kernel will not be
modified when generalizing to other distributions for the number of
crossovers (as long as they are in finite number
because of the stability of that law, cf. the explanation in the section on the absence of a steady-state solution) or by having
multiple
chromosomes, so we expect that even in these more realistic systems the long-term behavior will still see the genetic variance tend towards 0, though perhaps more slowly.
That key property should also hold even if the complex normal projection is not used, though of course it is not clear how to treat mathematically the dynamics of this system without introducing that projection.
Similarly, we do not expect the choice of form of selection to affect our conclusion, even though manifestly using truncation selection leads to unmanageable recursions~\cite{Kirkpatrick_2002,Santiago_1998,Turelli_Barton_1990,Turelli1994}. Note that, in our analysis, for convenience we selected gametes to have a given genetic value as a way of bypassing the difficulties generated by truncation selection, but the mathematics would have been hardly affected had we selected according to a Gaussian around that value.

Lastly, coming back to the controversy of the early 20th century between the two schools of thought, namely the biometric school and the Mendelian genetics school, let us note that within the biometricians, some expected that mixed heredity and its ``return to the mean'' under selection would drive traits to have vanishing variance in the absence of environmental noise, thus, in our language, they expected the variance of $G$ to converge to zero over generations. In that, they were right, but they also would also have predicted a bounded cumulative gain for $G$, and in that, they were wrong.

\section{Acknowledgments}
We acknowledge discussions with M. Falque, S. Majumdar, G. Schehr and B. Servin and we thank very warmly D. de Vienne, F. Kutle and A. Veber for their insights and advice. 
Most of all, we are indebted to J. Felsenstein for pointing us to the non-reducibility of the mixtures of complex normal distributions to complex normality when averaging over the crossover positions.

\section{Supplementary Material}
Supplementary material is available at PNAS Nexus online. It contains the scripts in R~\cite{Rsoft} used to generate the results ($N =$ 16383, 32767, 65535, 131071, 262143, 524287 using the selection parameter $\beta = 1.6$).

\section{Funding}
GQE and IPS2 benefit from the support of Saclay Plant Sciences-SPS (ANR-17-EUR-0007).

\section{Author contributions statement}
E.T. and O.C.M conceived the methodology and developed the associated computational scripts. E.T. conducted the  numerical runs. E.T. and O.C.M. analyzed and interpreted the results, and wrote the manuscript.

\section{Data availability}
All data used were generated by running the scripts in the Supplementary Material.



\bibliographystyle{abbrv}
\bibliography{references}


\vspace{2.0cm}
\begin{center}
\textbf{APPENDIX I: Fourier series of a recombined gamete}
\end{center}
Here we determine the coefficients of the Fourier series of a recombined gamete's profile as a function of those of its two constitutive gametes and of the angle $\psi$ specifying the position of the crossover $P_1$ as represented in Fig.\ref{Fig.crossovers}. We begin with the periodic step (Heaviside) function that is 1 in the intervals $]2n\pi,(2n+1)\pi[$ and vanishes otherwise. Its Fourier series representation is:
\begin{equation*}
H(\theta) = \frac{1}{2} + \frac{2}{\pi} \sum_{k=0}^{k=\infty} \frac{1}{(2k+1)} \sin{\left[(2k+1) \theta\right]} 
\end{equation*}
Let the two constitutive gametes in Fig.\ref{Fig.crossovers} have the profiles $x(\theta)$ and $x'(\theta)$. Each of these profiles has a Fourier series representation as given in Eq.\ref{Eq.fourier_series}; let the corresponding coefficients be $X_k$ and $X'_k$. The recombined gamete's profile is then $x''(\theta) = x(\theta)H(\theta-\psi) + x'(\theta)H(\theta -\psi - \pi)$. Since this is linear in $x$ and $x'$, the Fourier coefficients of a recombined gamete are linear in the Fourier coefficients of its constitutive gametes. Consider for instance the contribution of the first constitutive gamete using its series representation:
\begin{equation*}
x(\theta)H(\theta-\psi) = \left[ \sum_{-\infty}^{+\infty} e^{i k \theta} X_k \right] \left[ \frac{1}{2} + \frac{-i}{\pi} \sum_{-\infty}^{+\infty} \frac{e^{i (2k+1) (\theta - \psi)}}{2k+1} \right]
\end{equation*}
Carrying out this product and extracting the coefficient of $\exp(i k \theta)$ gives the $k$th Fourier coefficient for that first constitutive gamete's contribution to the recombined one:
\begin{equation*}
\frac{X_k}{2} + \frac{-i}{\pi} \sum_{-\infty}^{+\infty} \frac{X_{k-(2k'+1)}}{2k'+1} e^{-i (2k'+1) \psi}
\end{equation*}
The contribution from the second constitutive gamete is obtained similarly after the replacement of $X$ by $X'$ and of $\psi$ by $\psi + \pi$. Adding the contributions of both of these, we obtain that the $k$th Fourier coefficient of the series for the recombined gamete is:
\begin{eqnarray*}
X_k^{''} = && \frac{X_k + X_k^{'}}{2} 
+ \frac{-i}{\pi} \sum_{-\infty}^{+\infty}
\frac{X_{k-(2k'+1)}}{2k'+1} 
e^{-i (2k'+1) \psi} \\
&& + \frac{i}{\pi} \sum_{-\infty}^{+\infty}
\frac{X_{k-(2k'+1)}^{'}}{2k'+1} e^{-i (2k'+1) \psi}
\end{eqnarray*}
A consequence of this linear relation is that at all generations the distribution of Fourier coefficients 
-- conditional on the different $\psi$ angles -- 
remains multi-dimensional complex normal since their initial distribution is so.

\vspace{2.0cm}
\begin{center}
\textbf{APPENDIX II: Recursions for the variances of the Fourier coefficients}
\end{center}
As explained in the main part of the paper, because of the rotational invariance of the distribution of profiles ($\theta \to \theta + \delta$), the (complex) covariance matrix of the Fourier coefficients (Eq.\ref{Eq.complexNormal}) is diagonal. Since $X_k^{''}$ is a linear combination of independent random variables, its (complex) variance is just the sum of the complex variances of those variables. Let us first calculate the expectation of $\overline{X_k^{''}} X_k^{''}$:
\begin{eqnarray*}
E\left[ \overline{X_k^{''}} X_k^{''} \right] = && \frac{E\left[ \overline{X_k} X_k \right] + E\left[ \overline{X_k^{'}} X_k^{'} \right]}{4} \\
&& + \frac{1}{\pi^2} \sum_{-\infty}^{+\infty}
\frac{\left[ \overline{X_{k-(2k'+1)}} X_{k-(2k'+1)} \right]}{(2k'+1)^2} \\
&& + \frac{1}{\pi^2} \sum_{-\infty}^{+\infty}
\frac{\left[ \overline{X_{k-(2k'+1)}^{'}} X_{k-(2k'+1}^{'} \right]}{(2k'+1)^2}
\end{eqnarray*}
In the notation of the main part of the paper, this leads to the recursion when going from one generation to the next:
\begin{equation*}
\sigma_{k,g+1}^2 = \frac{\sigma_{k,g}^2}{2}
+ \frac{2}{\pi^2} \sum_{-\infty}^{+\infty}
\frac{\sigma_{k-(2k'+1),g}^2}{(2k'+1)^2}
\label{Eq.recursion_no_selection}
\end{equation*}
Using the diffusion interpretation presented in the main part of the paper, we see that half of the matter at each site is left in place while the other half is shared across other sites according to a convolution kernel decaying inversely with the square distance on the line of (odd) integers. This convolution kernel is normalized to 1 as it should be because of the result: 
\begin{equation*}
\frac{4}{\pi^2} \sum_{-\infty}^{+\infty}
\frac{1}{(2k'+1)^2} = 1
\end{equation*}
In the context of cycles where one first selects gametes having a given value for $G$, 
the $k=0$ Fourier component for all gametes are first set to that value before one proceeds with the mating. This then sets the corresponding variance to 0 and so, in the recursion formula given above, $\sigma_{k=0,g}^2$ must be replaced by 0, and then one recovers Eq.\ref{Eq.recursion}.

\end{document}